# PARISROC, a photomultiplier array readout chip

G. Martin-Chassard[*], S. Conforti, F. Dulucq, Mowafak El Berni, C. de La Taille, W. Wei [a]

*OMEGA/LAL/IN2P3, centre universitaire BP34 91898 ORSAY Cedex – France*

[a] *IHEP, Beijing, China*

**Abstract**

PARISROC is a complete read out chip, in AMS SiGe 0.35μm technology, for photomultipliers array. It is a front-end electronics ASIC which allows triggerless acquisition for the next generation of neutrino experiments. These detectors have place in megaton size water tanks and will require very large surface of photo-detection. An R & D program, funded by French national agency for research and called PMm2, proposes to segment the very large surface of photo-detection in macro pixels made of 16 photomultiplier tubes connected to an autonomous front-end electronics. The ASIC allows triggerless acquisition and only send out the relevant data by network to the central data storage. This data management reduces considerably the cost of these detectors. This paper describes the front-end electronics ASIC called PARISROC which integrates totally independents 16 channels with a variable gain and provides charge and time measurement with a 12-bit ADC and a 24-bits Counter.

## 1. Introduction

The PMm2 project [1] proposes an innovative electronics for array of photo-detectors used in high energy physics and astroparticle. The goal is to develop a macro pixel made of 16 small photomultiplier tubes connected to an autonomous front-end electronics, as shown in Fig.1, in order to segment very large surface of photo-detection.

This R&D [2] involves three French laboratories (LAL Orsay, LAPP Annecy, IPN Orsay) and PHOTONIS, a French photomultiplier tube maker. It is funded for three years by the French National Agency for Research (ANR) under the reference ANR-06-BLAN-0186.

The micro-electronics group (OMEGA) from the LAL at Orsay is in charge of the front-end electronics and has developed an ASIC called PARISROC (Photomultiplier ARray Integrated in Sige ReadOut Chip).

This paper describes the chip in detail and gives some measurement results.

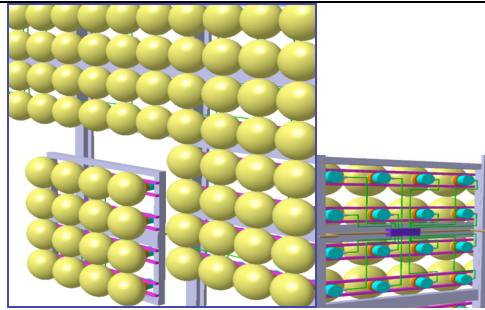

Figure 1 : A module with 16 PMT and a central readout ASIC

## 2. PARISROC architecture

### 2.1. requirements

The electronics must be autonomous, triggerless and with 16 independent channels. It must provide charge and time measurements. The charge measurement has to be efficient for a single photo-electron and keeps a good

- Corresponding author. Tel.: +33-164468318; fax: +33-164468934; e-mail: martin@lal.in2p3.fr.

linearity on a dynamic range up to 300 photo-electrons. The time measurement is done in two steps: a coarse measure with a 24 bit counter at 10 MHz and a fine measure on a 100 ns ramp to achieve a resolution of 1 ns. Because of the common high voltage supply, the preamplifier gain must be adjustable channel by channel in order to compensate the photomultiplier tube gain variation. Output data are serialized to be sent by only one communication wire.

*2.2. Global architecture*

The global architecture of the ASIC is shown in Figure 2. It is composed of 16 analogue channels managed by a common digital part.

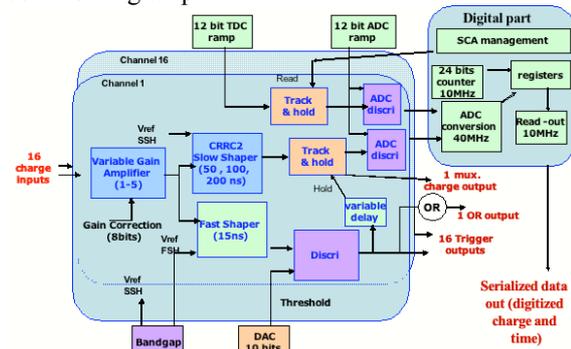

Figure 2: PARISROC global schematic

A bandgap block provides the common voltage references for fast and slow shapers. The threshold of the discriminators is common for the sixteen channels and given by a 10-bit DAC. The ramps for ADC and TDC are common for all the channels.

*2.3. One channel detail*

The detail of one analogue channel is given in Figure 3. Each channel is composed by a low noise preamplifier with variable and adjustable gain. The variable gain is common for all channels and can change from 8 to 1 on 4 bits. The gain is also adjustable channel by channel by a factor 4 on 8 bits.

The preamplifier is followed by a slow channel for the charge measurement and a fast channel for the trigger output. The slow channel includes a slow shaper followed by an analogue memory with a depth of 2 to provide a linear charge measurement up to 50 pC; this charge is converted by a 12-bit Wilkinson ADC. The fast channel is composed of a fast shaper (15ns) followed by 2 low offset discriminators to auto-trig down to 10 fC. The thresholds are loaded by 2 internal 10-bit DACs common for the 16 channels. The 2 discriminator outputs are multiplexed to provide only 16 trigger outputs. The fine time measurement is made by an analogue memory with depth of 2 which samples a 12-bit ramp as the same time of holding the slow shaper output. The analog memory is, then, converted by a 12-bit Wilkinson ADC.

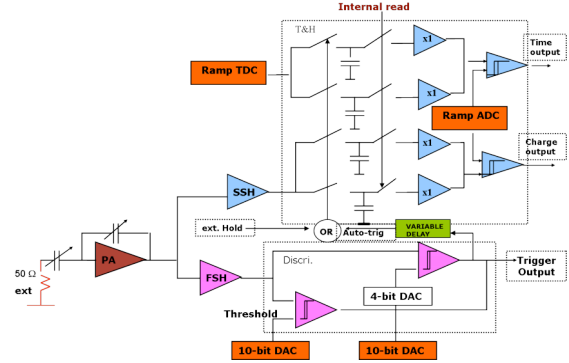

Figure 3: One channel schematic

*2.4. Digital part*

On overview of the digital part is given in Figure 4. The digital block manages the track and hold system like a FIFO and starts and stops all the counters [3]. All the data are serialized to be sent out. There are two clocks: one at 40 MHz for the analogue to digital conversion and for the track and hold management, the second at 10 MHz for timestamp and readout. The readout format is 52 bits: 4bits for channel number, 24 bits for timestamp, 12 bits for charge conversion and 12 bits for fine time conversion. The readout is selective: only the hit channels are read; so the maximum readout time will be 100 µs if all channels are hit.

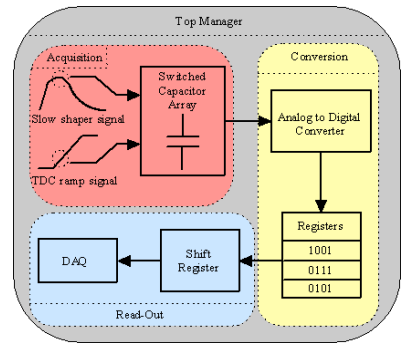

Figure 4: Digital part overview

*2.5. PARISROC layout*

The circuit has been designed in AMS SiGe 0.35µm technology [4]. The die (fig. 5) has a surface of 17 mm$^2$ (5mm X 3.4mm) and is packaged in a CQFP160 case.

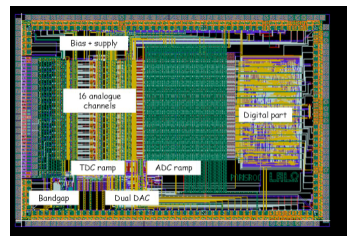

Figure 5: PARISROC layout

## 3. First measurement results

### 3.1. Test bench

A dedicated test board (Fig. 6) was developed to characterize the chip. It contains an FPGA to load slow control parameters and manage the ASIC outputs. It is driven by a USB and managed by a Labview program.

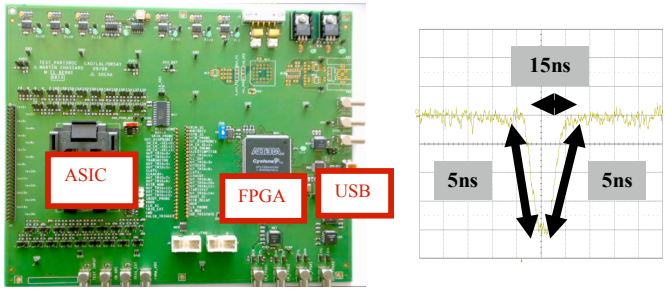

Figure 6 : Test board photography and input signal shape

All the measurements are done with an input signal made by a generator, as shown in fig. 6, to be like a PM tube signal. The voltage goes from 0 to 250mV to represent a charge up to 50 pC for a PM tube gain taken at $10^6$.

### 3.2. Trigger measurements

Noise at the output of the fast shaper is 2.5 mV in RMS (equivalent to 0.08 p.e.). One photo-electron signal gives 30mV at the fast shaper output with a rise time of 7 ns as shown in Fig. 7. The signal-to-noise ratio is more than 12. The fast shaper output is then sent to a discriminator for which the threshold is provided by an internal 10-bit DAC. The linearity of this DAC is measured to be 0.1%.

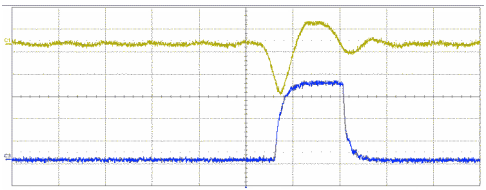

Figure 7 : fast shaper and discriminator outputs

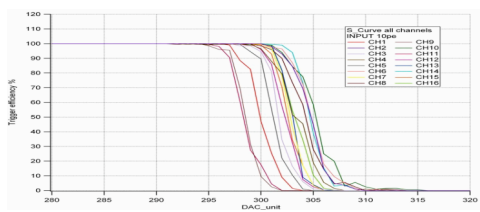

Figure 8 : Discriminator S-curves

The trigger efficiency versus threshold with an injected charge of 10 photo-electrons for all channels is given in the S-curves of fig.8. The spread of these curves is less than 7 LSB, equivalent to 15 mV or 0.5 photo-electron as the LSB of the DAC is 2 mV (0.07 pe). The noise, given by the slope of the curve, is less than 0.1 photo-electron.

### 3.3. Charge measurements

Figure 9 shows the maximum preamplifier output versus inverse of the feedback capacitance. This plot gives the linearity of the preamplifier gain variation (about 2%).

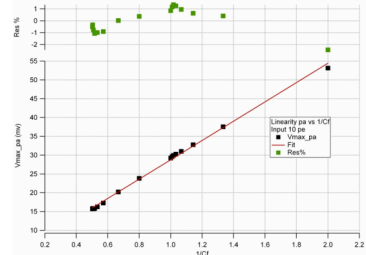

Figure 9 : preamplifier gain linearity

The slow shaper has a good linearity (around 1%) but has more noise than expected due to an extra low noise frequency which is now under investigation to be reduced.

When there is a trigger, it is delayed to hold the shaper signal at its maximum as shown in Fig. 10.

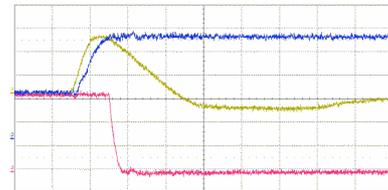

Figure 10 : slow shaper, trigger delayed and shaper held signals

We measure the linearity of the internal 12-bits ADC : results are shown in Fig. 11 and gives a non-linearity less than plus and minus one LSB with an RMS of 0.3 LSB.

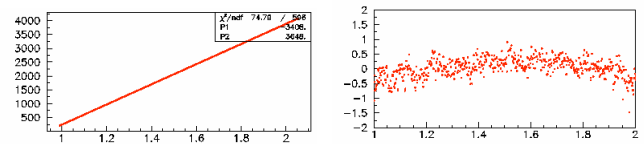

Figure 11 : 12-bit ADC linearity and residuals

## 4. Conclusion

The first measurements of PARISROC are encouraging except for the slow shaper noise but a lot of measurements are left in order to characterize a so complex chip.

The module of PMm2 with 16 photomultiplier tubes and a central ASIC will be built at the end of this year.